\documentclass[showpacs,preprintnumbers,amsmath,amssymb,superscriptaddress,nofootinbib,english]{revtex4}

\usepackage{epstopdf}
\usepackage{graphicx}
\usepackage{dcolumn}
\usepackage{bm}
\usepackage{epsfig}
\usepackage{graphicx}
\usepackage{hyperref}
\usepackage[usenames]{color}
\usepackage{url}

\hypersetup{
    colorlinks=true,
    linkcolor=green,
    citecolor=blue,
}

\newcommand{\remove}[1]{}

\def\be{\begin{equation}}
\def\ee{\end{equation}}
\def\ba{\begin{eqnarray}}
\def\ea{\end{eqnarray}}

\begin{document}

\title{The simplest extension of Starobinsky inflation}
\author{Carsten van de Bruck}
\email[Email address: ]{C.vandeBruck@sheffield.ac.uk}
\affiliation{Consortium for Fundamental Physics, School of Mathematics and Statistics, University of Sheffield, Hounsfield Road, Sheffield, S3 7RH, United Kingdom}

\author{Laura E. Paduraru}
\email[Email address: ]{LEPaduraru1@sheffield.ac.uk}
\affiliation{Consortium for Fundamental Physics, School of Mathematics and Statistics, University of Sheffield, Hounsfield Road, Sheffield, S3 7RH, United Kingdom}

\date{\today}

\begin{abstract}
We consider the simplest extension to the Starobinsky model, by allowing an extra scalar field to help drive inflation. We perform our analysis in the Einstein frame and calculate the power spectra at the end of inflation to second order in the slow--roll parameters. We find that the model gives predictions in great agreement with the current Planck data without the need for fine-tuning. Our results encourage current efforts to embed the model in a supergravity setting. 
\end{abstract}

\maketitle

\section{Introduction}

The theory of inflation, a time of accelerated expansion of the early universe, was first proposed in the 1980's and since then a myriad of models have been studied to explain the mechanism underlying it. The issue with inflation is that it is a phenomenological construct, which needs to be embedded into a fundamental theory. In the very first model the period of inflation was driven by quantum corrections to the Einstein--Hilbert Lagrangian \cite{Starobinsky:1980te}. In its simplest and most studied version, the Einstein--Hilbert action includes an additional term quadratic in the Ricci--scalar $R$ and is usually called Starobinsky--inflation or also $R^2$ inflation. Later models of inflation are based on the dynamics of scalar fields \cite{Guth:1980zm,Linde:1981mu,Albrecht:1982wi}. We refer to \cite{Martin:2014vha} and \cite{Mazumdar:2010sa} for reviews and list of references on inflationary models. Another way of generating inflation is by considering  further modifications of General Relativity. Models include Gauss--Bonnet gravity \cite{Satoh:2010ep, Guo:2010jr, Koh:2014bka, Neupane:2014zxa, Kanti:2015pda} and higher--order--polynomial corrections \cite{Huang:2013hsb} (see also \cite{Artymowski:2015mva} for a recent discussion on higher--order corrections in Starobinsky inflation).

There has been revived interest in Starobinsky inflation after Planck 2013 results \cite{Ade:2013zuv}, which place it in a favourable light with respect to cosmological observables. This model of inflation is remarkably consistent with current cosmological data, in particular with measurements of the anisotropies of the CMB. It predicts a spectral index $n_s \approx 0.96$ with little spectral running and a small amount of gravitational waves. 

The idea of generating $R^2$--inflation from a more fundamental theory has come into focus. Similar effort to what  was done in finding a realisation of Higgs inflation in supergravity, either to formulate the theory in the Jordan frame (see e.g. \cite{Ferrara:2010yw, Nakayama:2010ga}) or in the Einstein frame \cite{BenDayan:2010yz}, has been invested in Starobinsky inflation. Several papers have tried to embed the model into a fundamental framework, such as supergravity (see e.g. \cite{Ellis:2013xoa,Ketov:2012jt,Ferrara:2013rsa,Kallosh:2013lkr,Ferrara:2014ima,Kounnas:2014gda,Ellis:2014gxa,Ellis:2014opa, Diamandis:2014vxa}). Motivation of this work was also to extend the model in such a way that a large tensor-to-scalar ratio can be obtained.
In the context of Supergravity, the inflaton fields are components of chiral multiplets (or are combinations thereof) with the vector components are not playing an important role and the universe is assumed to behave like a Friedmann--Robertson--Walker (FRW) universe on large scales. Some of the cases considered are multi--field models, in the sense that not only one field contributes to the dynamics of inflation.

The question can be asked, whether Starobinsky inflation preserve its attractive properties in the presence of interacting matter fields. There is no reason to assume that gravity is the only driving force behind inflation, so in our setup we consider an extension of the Starobinsky model by including a scalar field in the matter sector. These have been studied in the past \cite{Gottlober:1990um} to obtain a period of double inflation. In our paper, we study the system in the Einstein frame, calculating the amplitude of the curvature perturbation at the end of inflation, the spectral index and the tensor-to-scalar ratio. The Einstein frame analysis allows us to use the formalism presented in \cite{vandeBruck:2014ata} to calculate the power spectra for a large number of model degrees of freedom and initial conditions without using the full field equations. In \cite{vandeBruck:2014ata}, the power spectra were calculated up to second order in the slow--roll parameter. 

Our motivation is two-fold: firstly, we want to explore the robustness of Starobinsky-type inflation in the presence of matter fields, since these fields are not necessarily expected to be dynamically insignificant. If these fields contribute to the dynamics of the very early universe, then predictions of the Starobinsky model will potentially be altered and our aim is to quantify this further. Secondly, embedding the Starbinsky model in a supergravity framework motivates the existence of more than degree of freedom, which are usually ignored. Thus, we take a phenomenological approach, allowing two degrees of freedom to evolve. The theory we consider has two free parameters, as we will discuss below.  

The paper is organised as follows: In the next Section we present the theoretical setup and describe our numerical method. In Section III we discuss our findings. We conclude in Section IV.

\section{Theoretical setup}
The theory we consider is the simplest extension of Starobinsky's original model of inflation we can imagine. It includes an $R^2$--term in the Einstein-Hilbert action as well as a massive scalar field $\chi$ in the matter sector. The full action reads 

\begin{equation}
\begin{split}
S &= \int d^{4}x \sqrt{-g} \bigg[ \frac{R}{2 \kappa} + \frac{\mu}{2} R^2 \bigg] + \int d^{4}x \sqrt{-g}\bigg[ - \frac{1}{2} g^{\mu \nu}\partial_{\mu}\chi \partial_{\nu}\chi - \frac{1}{2} m_\chi^2 \chi^{2}\bigg]
\end{split}
\end{equation}
In this equation, $\kappa = M_{\rm Pl}^{-2}$, where $ M_{\rm Pl}^{-2}$ is the reduced Planck mass. The parameter $\mu$ has units [mass]$^{-2}$. 
We will perform the analysis in the Einstein--frame, which can be obtained by a conformal transformation. First, we rewrite the gravitational sector of the action above as 
\begin{equation}
\label{eq:reform}
S^{\prime}  _{G}= \int d^{4}x \sqrt{-g} \bigg[ \frac{R}{2 \kappa}(1+2 \kappa \mu \phi) - \frac{\mu \phi^{2}}{2}\bigg] 
\end{equation}
The equation of motion for $\phi$ gives $\phi = R$, thus the two action are equivalent. Now considering the conformal transformation 
\begin{equation}
\tilde g_{\mu\nu} = \Omega^{2} g_{\mu\nu} 
\end{equation}
with 
\begin{equation}
\Omega^{2} = 1+2\kappa \mu \phi, 
\end{equation}
we obtain from eq. (\ref{eq:reform}) 

\begin{equation}
\begin{split}\label{einsteinframeaction}
S_E = \int d^{4}x \sqrt{-\tilde g} \bigg[& \frac{ \tilde R }{2 \kappa}  - \frac{\tilde g^{\mu \nu}}{2} (\tilde \partial _{\mu} \psi)(\tilde \partial_{\nu} \psi)  -\frac{1}{2} \tilde g^{\mu \nu}e^{-2\alpha \psi}(\tilde\partial_{\mu}\chi)(\tilde \partial_{\nu}\chi) - V \bigg]\\
\end{split}
\end{equation}
with 
\begin{equation}
V =     \frac{(e^{2\alpha \psi}-1)^{2}}{ 8 e ^{4\alpha \psi}\kappa^2 \mu}+ \frac{1}{2} m_\chi^2 e^{-4\alpha \psi} \chi^{2}
\end{equation}
and $\alpha = \frac{\kappa}{\sqrt{6}}$ and where we have defined the canonical normalised field $e^{2\alpha\psi} = 1+2\kappa \mu \psi$. We now define the mass of the scalaron $\psi$ as \cite{Gottlober:1990um}: 

\begin{equation}
m_\psi^2 = \frac{1}{6\kappa\mu}~.
\end{equation}
In the following, we will work in natural units, i.e. we set $\kappa = 1$. This type of theory has been considered in the past in \cite{Gottlober:1990um}, working in the Jordan frame. The model discussed in \cite{Ellis:2014gxa}, with the choice $\omega = 1$, $m = m_\chi$ and $a^2 = 1/12$, is a special case of our model, where the fields have the same mass. Our model allows the two fields to have different masses $m_\psi$ and $m_\chi$.

Following the literature, we define $b(\psi) = -2\alpha\psi$ and derive the equations of motions for the fields in an expanding Robertson--Walker space-time (the dot denotes the derivative with respect to cosmic time):
\begin{eqnarray}
\ddot{\psi}+3H\dot{\psi}+V_\psi &=& b_\psi e^{2b}\dot{\chi}^2, \\
\ddot{\chi}+(3H+2b_\psi \dot{\psi})\dot{\chi}+e^{-2b}V_\chi &=& 0, 
\end{eqnarray}
Einstein's field equation give
\begin{eqnarray}
\dot{H} &=& -\frac{1}{2M_P^2}\left[\dot{\psi}^2+e^{2b}\dot{\chi}^2 \right]\quad \text{and}\\
H^2 &=& \frac{1}{3M_P^2}\left[\frac{\dot{\psi}^2}{2} + \frac{e^{2b}}{2}\dot{\chi}^2 + V \right].
\end{eqnarray} 

To study the perturbations produced during inflation, we will not work with the fields $\psi$ and $\chi$ but perform a field--rotation using the degree of freedom 
along the field trajectory (denoted $\sigma$) and the degree of freedom orthogonal to it (denoted $s$). The fields are defined by 
\begin{eqnarray}
d\sigma &=& \cos\theta d\psi + \sin\theta e^b d \chi \\
ds &=& e^b\cos\theta d\chi - \sin\theta d\psi,
\end{eqnarray}
with 
\begin{eqnarray}
\cos\theta &=& \frac{\dot\psi}{\sqrt{\dot\psi^2 + e^{2b}\dot\chi^2}} ~, \nonumber \\
\sin\theta &=& \frac{e^b \dot\chi}{\sqrt{\dot\psi^2 + e^{2b}\dot\chi^2}}~.
\end{eqnarray}

Cosmological perturbations in this system have been systematically studied in the past, see e.g. \cite{DiMarco:2002eb,Lalak:2007vi,vandeBruck:2014ata}. 
In \cite{vandeBruck:2014ata}, a formalism was developed, which allows to calculate the power spectra to second order in the slow--roll parameter. Instead of integrating the full perturbation equations, we will use this formalism to calculate the power spectrum. The method consists of two steps: Firstly, we evaluate the power spectrum at horizon crossing (the $*$ denotes the time of horizon crossing and the slow--roll parameter are defined with respect to $\sigma$ and $s$):
\begin{equation}
\begin{split}
\mathcal{P}_{\mathcal{R}*} &= \frac{H^2_*}{8\pi^2\epsilon_*} (1-2\epsilon_*-11\epsilon_*^2+4\epsilon_*\eta_{\sigma\sigma*}+4\epsilon_*\xi_{1*} s^2_{\theta*} c_{\theta*})(1+k^2\tau^2) \times \\
& \quad\quad\quad\left[1+ \frac{2}{3}(3\epsilon_*+20\epsilon_*^2-8\epsilon_*\eta_{\sigma\sigma*}-8\epsilon_*\xi_{1*} s^2_{\theta*} c_{\theta}* -A_{Q*})f(x) \right.\\
& \quad\quad\quad\quad \left.\ + \left(\epsilon_*^2+\frac{A_{Q*}^2+B_{Q*}^2}{9}-\frac{2\epsilon_* A_{Q*}}{3} \right) g(x) \right],
\end{split}
\end{equation}
\begin{equation}
\begin{split}
\mathcal{P}_{\mathcal{S}*} &= \frac{H^2_*}{8\pi^2\epsilon_*} (1-2\epsilon_*-11\epsilon_*^2+4\epsilon_*\eta_{\sigma\sigma*}+4\epsilon_*\xi_{1*} s^2_{\theta*} c_{\theta*})(1+k^2\tau^2) \times \\
& \quad\quad\quad\left[1+ \frac{2}{3}(3\epsilon_*+20\epsilon_*^2-8\epsilon_*\eta_{\sigma\sigma*}-8\epsilon_*\xi_{1*} s^2_{\theta*} c_{\theta}* -D_{Q*})f(x) \right.\\
& \quad\quad\quad\quad \left.\ + \left(\epsilon_*^2+\frac{D_{Q*}^2+B_{Q*}^2}{9}-\frac{2\epsilon_* D_{Q*}}{3} \right) g(x) \right],
\end{split}
\end{equation}
where 

\begin{eqnarray}
\begin{split}
A_Q &= 3\eta_{\sigma\sigma}-6\epsilon+3\xi_1 s^2_\theta c_\theta + 10\epsilon\eta_{\sigma\sigma}-18\epsilon^2+ 11\epsilon\xi_1 s^2_\theta c_\theta\\ & \quad \quad -\eta_{\sigma\sigma}\xi_1 s^2_\theta c_\theta + \xi_1^2 s^4_\theta - \eta_{\sigma s}\xi_1 s_\theta(1+c_\theta^2),\label{eq:Aeq}\\
B_Q &= 3\eta_{\sigma s} - 3\xi_1s_\theta^3 + 8\epsilon\eta_{\sigma s} - 9\epsilon\xi_1 s_\theta^3 + \eta_{\sigma\sigma}\xi_1 s_\theta^3 - \eta_{\sigma s}\xi_1 c_\theta^3+\xi_1^2s_\theta^3 c_\theta,\label{eq:Beq}\\
C_Q &= 3\eta_{\sigma s} - 3\xi_1s_\theta^3 + 8\epsilon\eta_{\sigma s} - 9\epsilon\xi_1 s_\theta^3 + \eta_{\sigma\sigma}\xi_1 s_\theta^3 - \eta_{\sigma s}\xi_1 c_\theta^3+\xi_1^2s_\theta^3 c_\theta,\label{eq:Ceq}\\
D_Q &= 3\eta_{ss} - 3\xi_1 c_\theta(1+s_\theta^2)+6\epsilon\eta_{ss} - 7\epsilon\xi_1 c_\theta(1+s_\theta^2)+ \eta_{\sigma\sigma}\xi_1 c_\theta(1+s_\theta^2)\\
& \quad\quad +\eta_{\sigma s}\xi_1 s_\theta c_\theta^2 + \xi_1^2(s_\theta^4-c_\theta^2) .\label{eq:Deq}
\end{split}
\end{eqnarray}
\begin{align}
f(x) &= 2-\gamma-\ln{2} -\ln{x},\\
6g(x)&= 16+3\pi^2-44\gamma+12\gamma^2+24\gamma\ln{2}-44\ln{2}+12\ln^2{2} \notag\\
&\quad\quad + 12\ln^2{x} -44\ln{x}+24\gamma\ln{x}+24\ln{x}\ln{2}. \label{eq:6gxeq}
\end{align}

The next step is to evaluate the power spectrum at the end of inflation. Because of the presence of isocurvature perturbations, the power spectrum of the curvature perturbation evolves on superhorizon scales. Defining 

\begin{equation}
\begin{split}
A &= \left(2\epsilon - \eta_{\sigma\sigma}-\xi_1 s_\theta^2 c_\theta-\frac{\eta_{\sigma s}^2}{3} -\frac{4\epsilon^2}{3}-\frac{\eta_{\sigma\sigma}^2}{3} +\frac{5\epsilon\eta_{\sigma\sigma}}{3} -\frac{2\xi_1^2 s^2_\theta c^2_\theta}{3}+ \frac{\xi_2 s^2_\theta c^2_\theta}{3}\right.\\
& \quad \left.  -\frac{4\eta_{\sigma\sigma}\xi_1 s^2_\theta c_\theta}{3}- \frac{4\eta_{\sigma s}\xi_1 s_\theta c_\theta^2}{3}+\frac{4\epsilon\xi_1 s^2_\theta c_\theta}{3} -\frac{\alpha_{\sigma\sigma\sigma}}{3}\right),
\end{split}
\end{equation}
\begin{equation}
\begin{split}
B &= \left(-2\eta_{\sigma s} +2\xi_1 s^3_\theta + 2\epsilon\eta_{\sigma s}-\frac{2\eta_{\sigma\sigma}\eta_{\sigma s}}{3}-\frac{2\eta_{ss}\eta_{\sigma s}}{3}+\frac{4\eta_{\sigma\sigma}\xi_1 s^3_\theta}{3}- \frac{4\epsilon\xi_1 s^3_\theta}{3}\right.\\
& \quad \left. - \frac{4\eta_{ss}\xi_1s_\theta c^2_\theta}{3}+ \frac{4\xi_1^2s^3_\theta c_\theta}{3}- \frac{2\alpha_{\sigma\sigma s}}{3}\right),
\end{split}
\end{equation}
\begin{equation}
\begin{split}
D &= \left(-\eta_{ss}+\xi_1c_\theta(1+s_\theta^2)-\frac{\eta_{\sigma s}^2}{3}-\frac{\eta_{ss}^2}{3}+\frac{\epsilon\eta_{ss}}{3}-\frac{\alpha_{\sigma ss}}{3} + \frac{4\eta_{\sigma s}\xi_1 s_\theta^3}{3} \right. \\
&\quad \left. - \frac{4\xi_1^2 s_\theta^4}{3} + \frac{4\eta_{ss}\xi_1 c_\theta s_\theta^2}{3}\right),
\end{split}
\end{equation}\label{eq:Geq}
the final power spectra are given by 
\begin{eqnarray}
\mathcal{P_R}(N) &=& \mathcal{P_R}_* \left[ 1 + \left(\displaystyle \int^{N}_{N_*} B(N'')e^{\int^{N}_{N_*}\gamma(N')dN'} dN'' \right)^2 \right. \nonumber  \\
 &-& \left. 2\eta_{\sigma s} f(-k \tau_*) \displaystyle \int^{N}_{N_*} B(N'') e^{\int^{N}_{N_*}\gamma(N')dN'} dN''  
 \right] ,\\
\mathcal{P_S}(N) &=& \mathcal{P_S}_*e^{2\int^{N}_{N_*}\gamma(N')dN'},
\end{eqnarray}
where the '$_*$' denotes the value of the power spectra at horizon crossing and $\gamma = D - A$.

Finally, to calculate the power spectrum of tensor perturbations $\mathcal{P}_{\mathcal{T}}$ we use the slow-roll approximation (see e.g. \cite{PU}):
\begin{equation}
\label{eq:Pt}
\mathcal{P}_{\mathcal{T}} = \frac{16}{\pi} \left[ 1 - 2(\gamma + \ln 2 - 1) \epsilon \right] \frac{H^{2}}{M_{\rm Pl}^{2}}.
\end{equation}

\section{Numerical Results}
The model we consider in this paper has two degrees of freedom ($\psi$ and $\chi$) with two free parameters ($m_\psi$ and $m_\chi$). The initial conditions are specified by the initial values of the fields and their derivatives. We start our fields at zero velocity and make sure that inflation lasts long enough, so that  the fields are in the slow--roll regime at the time the observable scales leave the horizon. We address the following questions: firstly, do the initial conditions of the fields have an significant influence on the observables and secondly, what are the restrictions on the masses of the fields in light of the latest {\it Planck} results? 

\begin{figure}[hb]
\begin{center}
\vspace*{-1.5cm}
\includegraphics[width = 0.5\linewidth,height = 0.424\textheight]{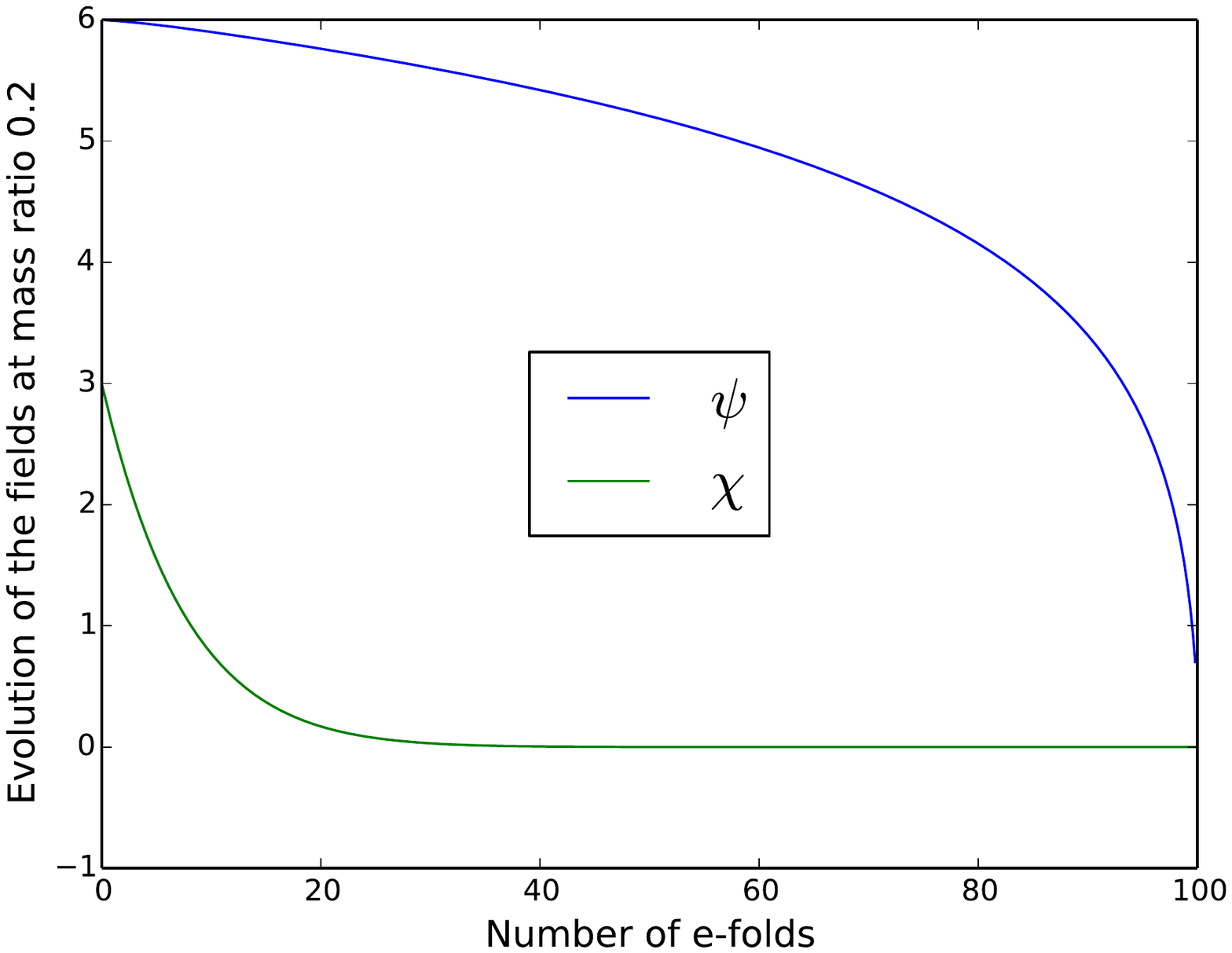}~~~~\includegraphics[width = 0.5\linewidth,height = 0.424\textheight]{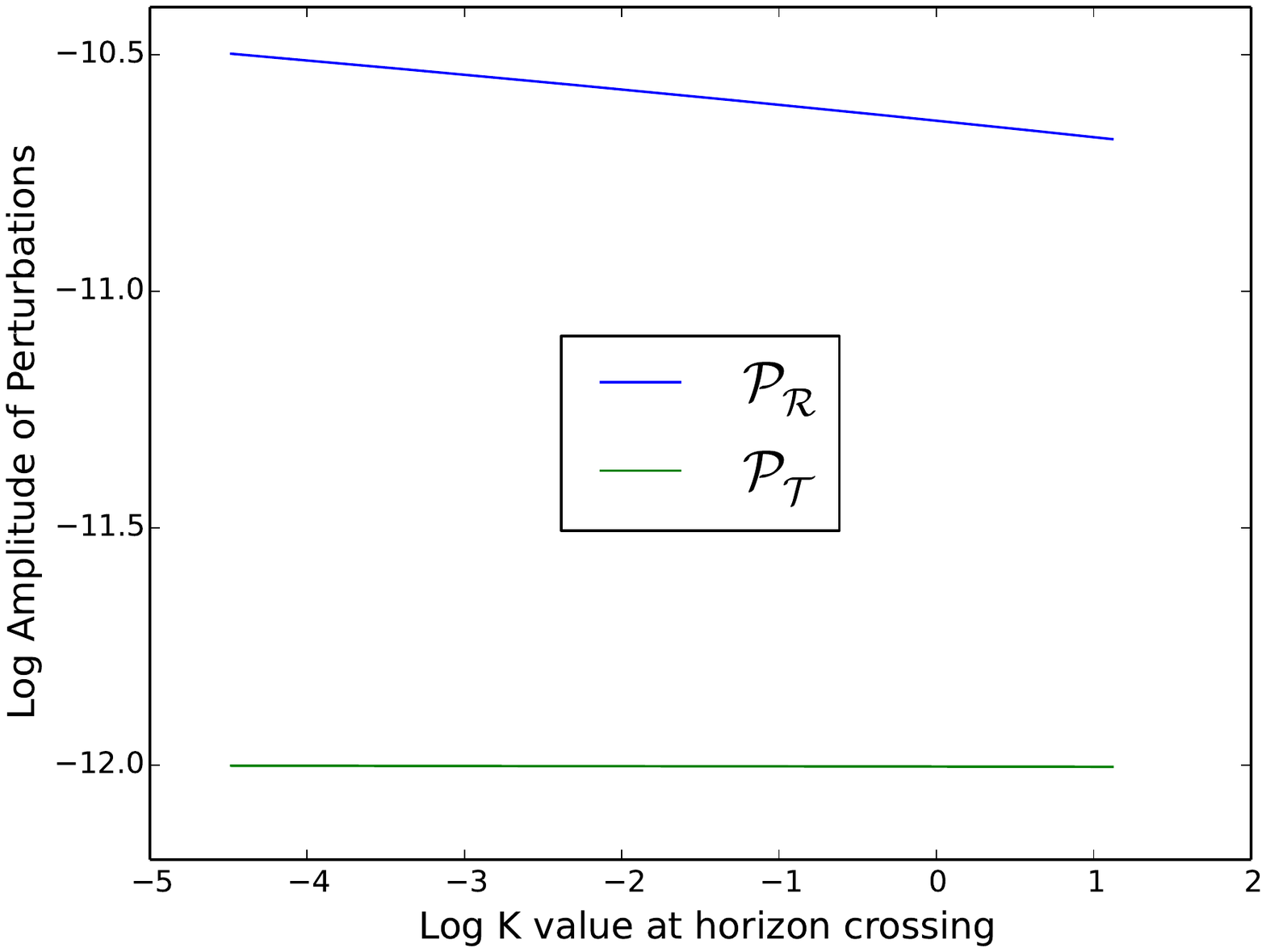}\\
\vspace*{-4cm}
\includegraphics[width = 0.5\linewidth,height = 0.424\textheight]{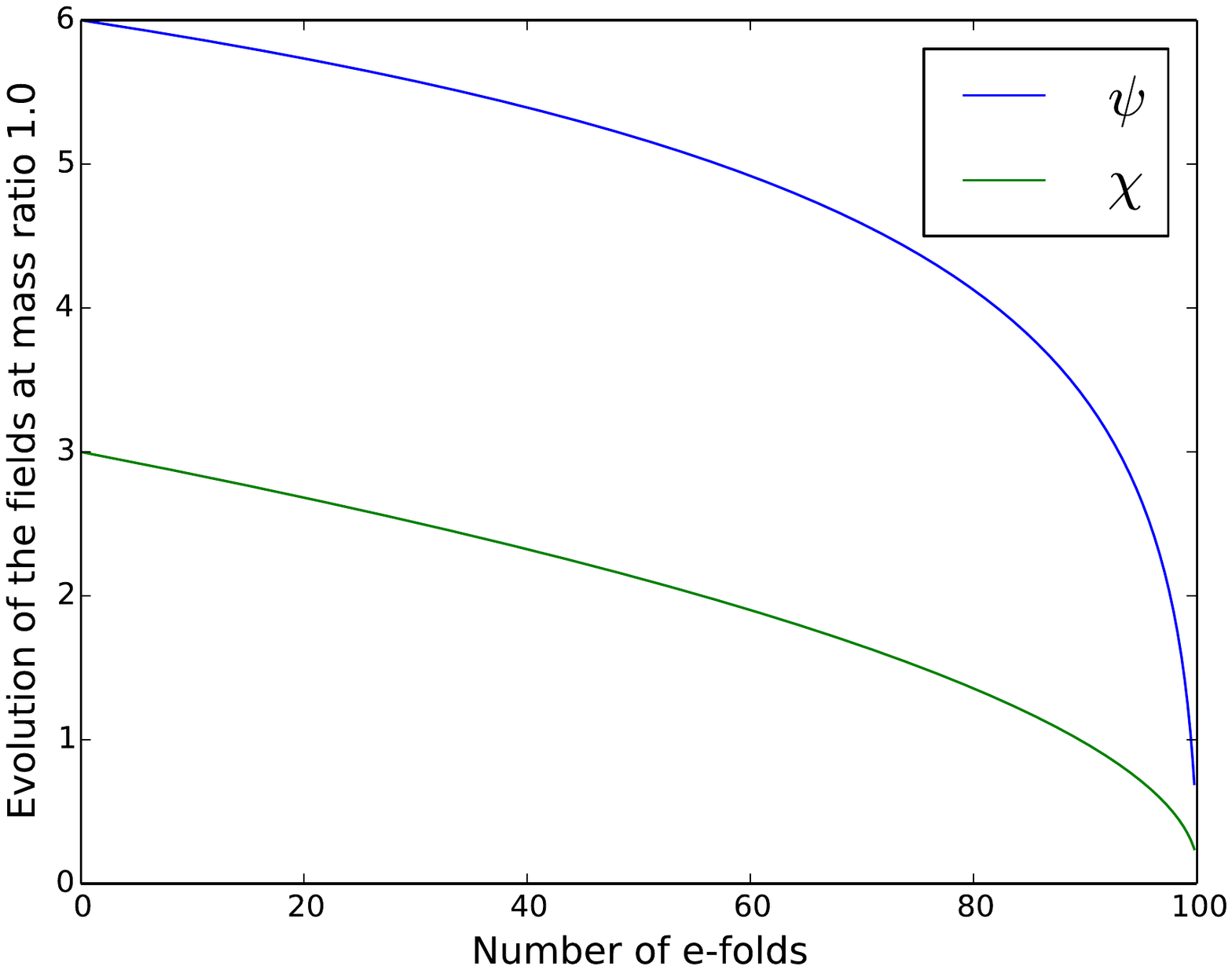}~~~~\includegraphics[width = 0.5\linewidth,height = 0.424\textheight]{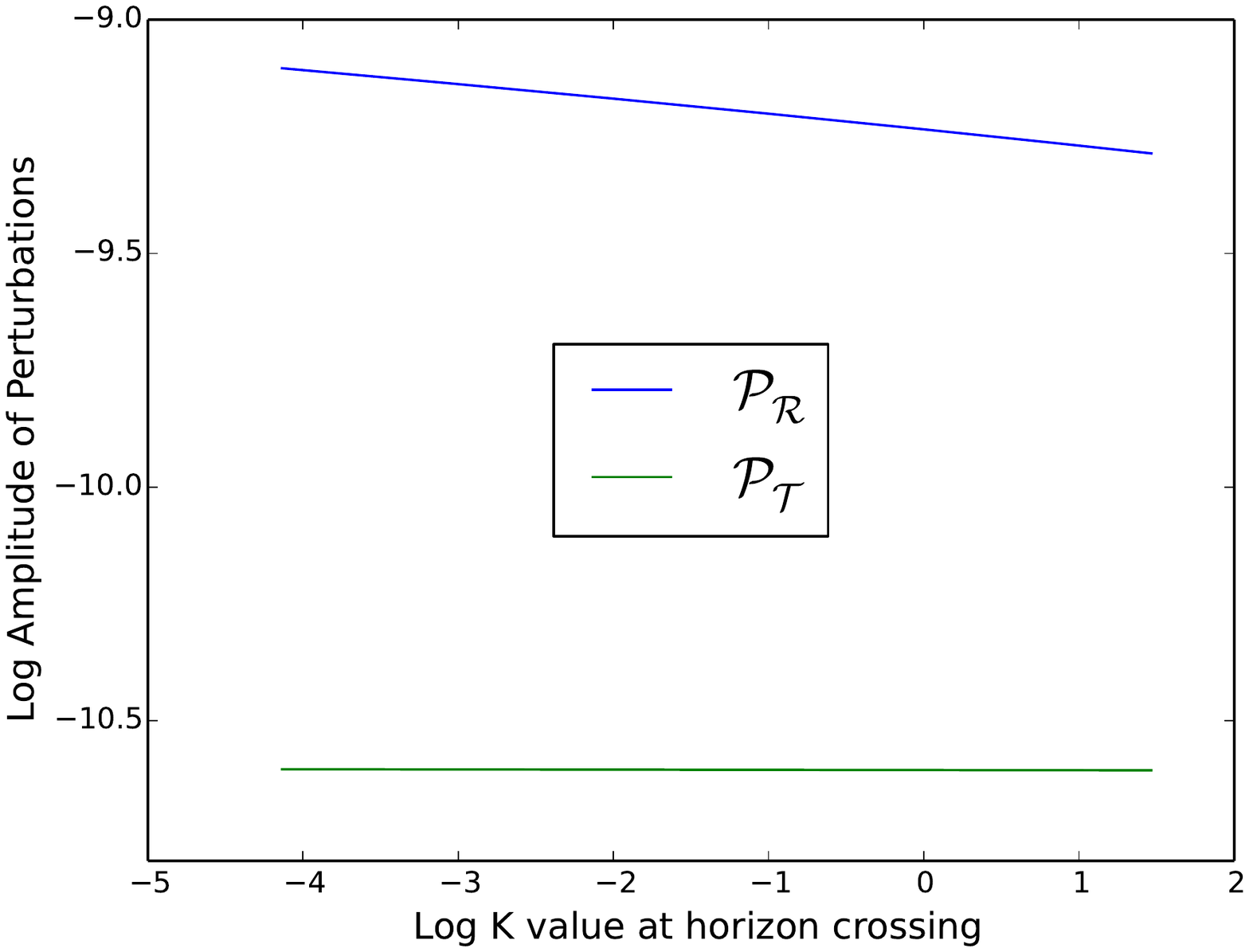}\\
\vspace*{-4cm}
\includegraphics[width = 0.5\linewidth,height = 0.424\textheight]{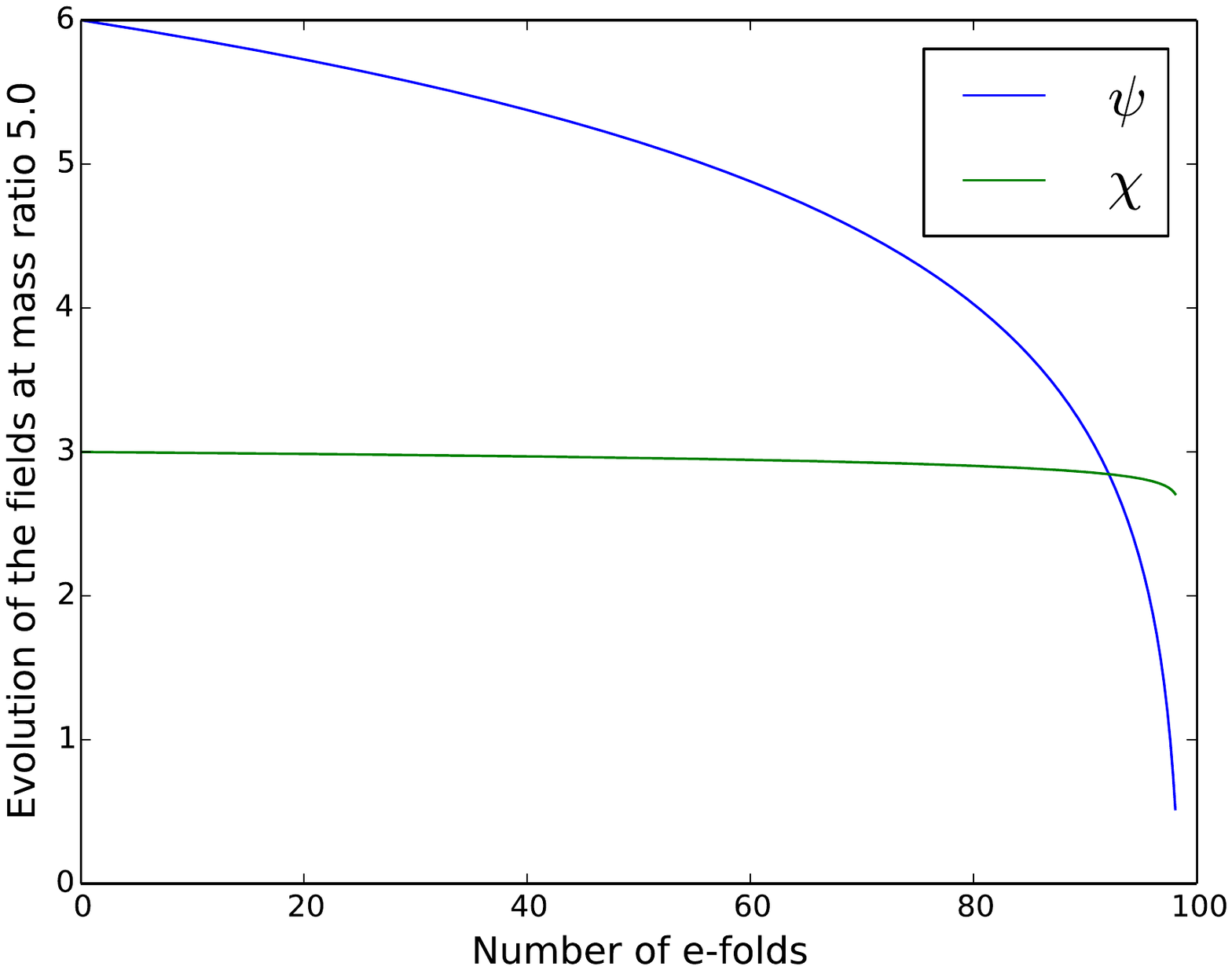}~~~~\includegraphics[width = 0.5\linewidth,height = 0.424\textheight]{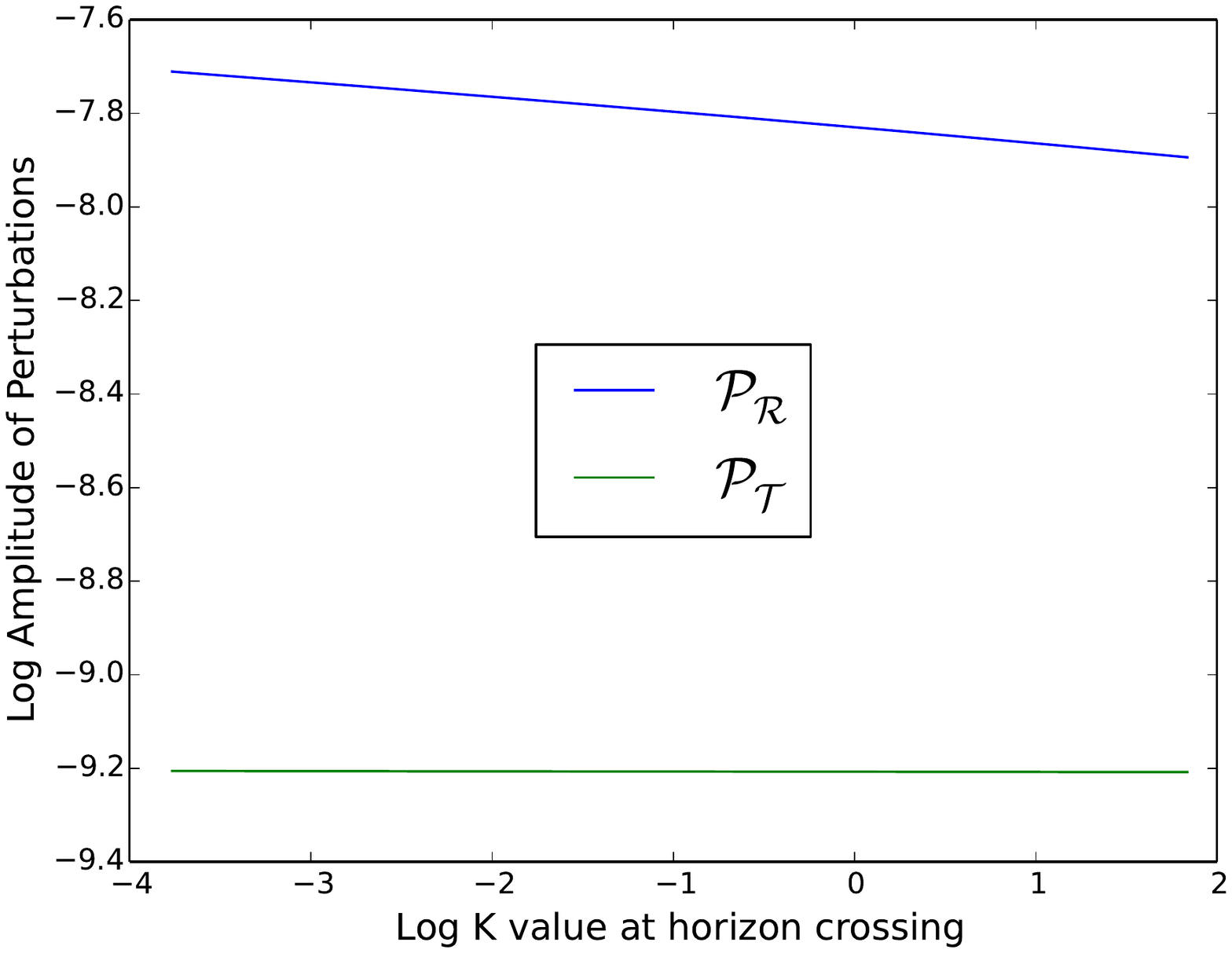}\\
\end{center}
\vspace*{-2cm}
\caption{Field trajectories and corresponding power spectra for runs with the same initial conditions and mass ratios $R_{m} = 0.2$, $R_{m} = 1.0$ and $R_{m} = 5.0$ respectively. The spectral indices for the curvature power spectrum is too large in these plots, the trajectories shown are for illustrative purposes only.}
\label{fig:background-examples}
\end{figure}

Using the formalism presented in the last section, we evaluate the power spectrum and calculate the amplitude $A_S$ of the power spectra at the pivot point $k_{\rm pivot} = 0.05h^{-1}$Mpc$^{-1}$, the spectral index $n_s$ and the tensor-to-scalar ratio $r$. To test the robustness of the predictions of the model, we run a large number of different initial conditions for the fields $\psi$ and $\chi$ for the mass ratio $R_{m}= m_{\psi}/m_{\chi} $ between 0.1 and 5. We find that the variation in the spectral index and the tensor--to scalar ratio is $< 1\%$. The tensor--to--scalar ratio $r$ is small, ranging from $r\approx 0.035$ to $r \approx 0.07$, with the details depending on $m_\psi$ and $m_\chi$. The variation in $A_S$ can be as large as $10\%$, for given combinations of $m_{\chi}$ and $R_{m}$, however that seems to be the exception rather than the rule. For the majority of regions in parameter space, the variation in $A_S$ is $<1\%$. The running of the spectral index is of order $10^{-4}$ or smaller and compatible with current observations.

In Figure \ref{fig:background-examples} we illustrate the three types of trajectories that the fields can follow, as they are approaching the global minimum at the end of inflation and the associated power spectra. The figure shows that varying the mass ratio $R_m$ changes the behaviour of the fields. The heavier $\psi$ becomes, the more the model behaves like double inflation, where the heavy field drives inflation first and then the second field triggers a second period of inflation. In the opposite case, where $\chi$ is heavier, it quickly approaches zero, leaving inflation to be driven by the $\psi$ field. When calculating the power spectra, we compared the second--order formalism presented in the last section to the results using a first--order formalism and found very good agreement. This confirms that for the purpose of this paper a full integration of the equations is not necessary. 

\begin{figure}[hb]
\begin{center}
\includegraphics[width=0.5\textwidth]{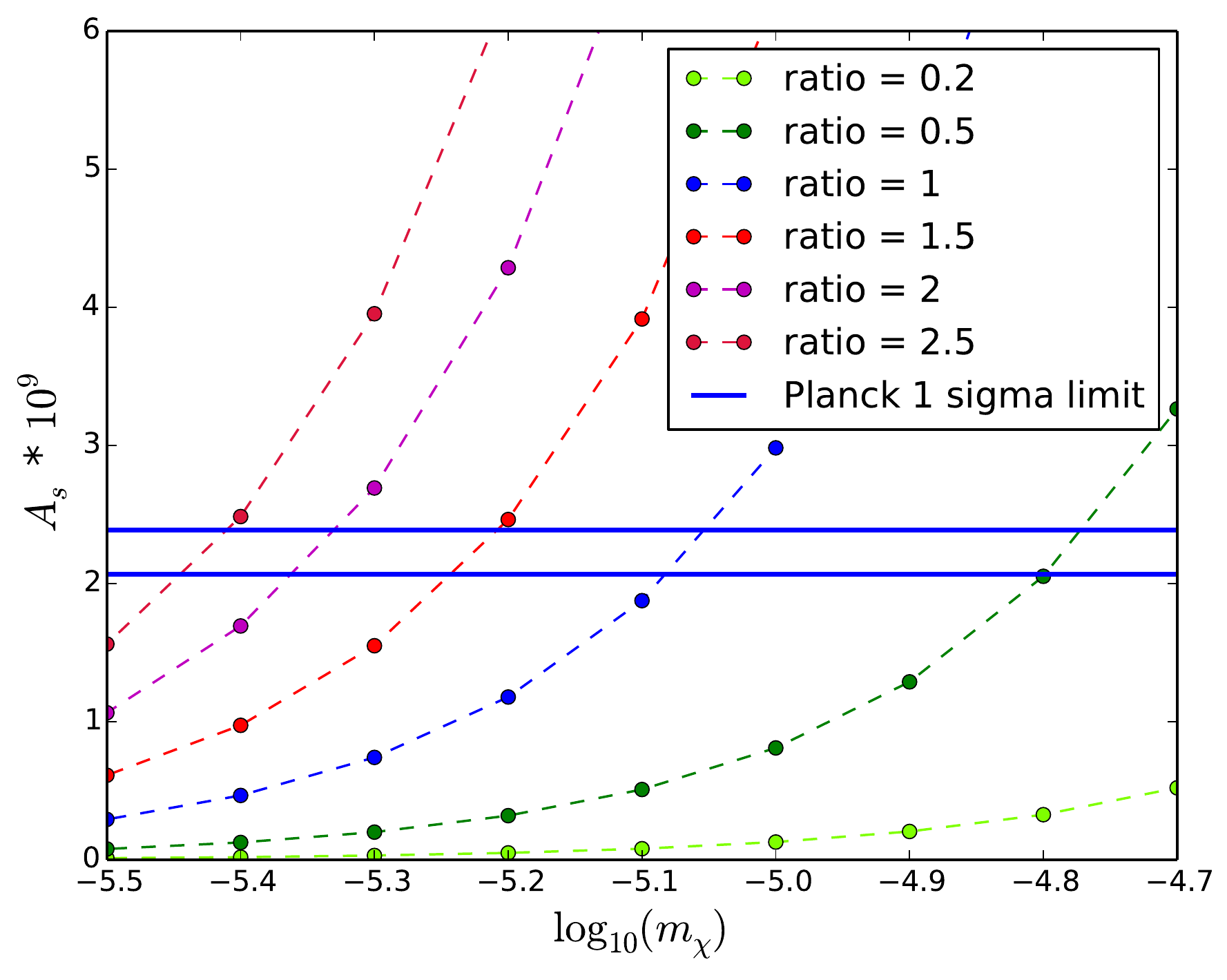}
\includegraphics[width=0.5\textwidth]{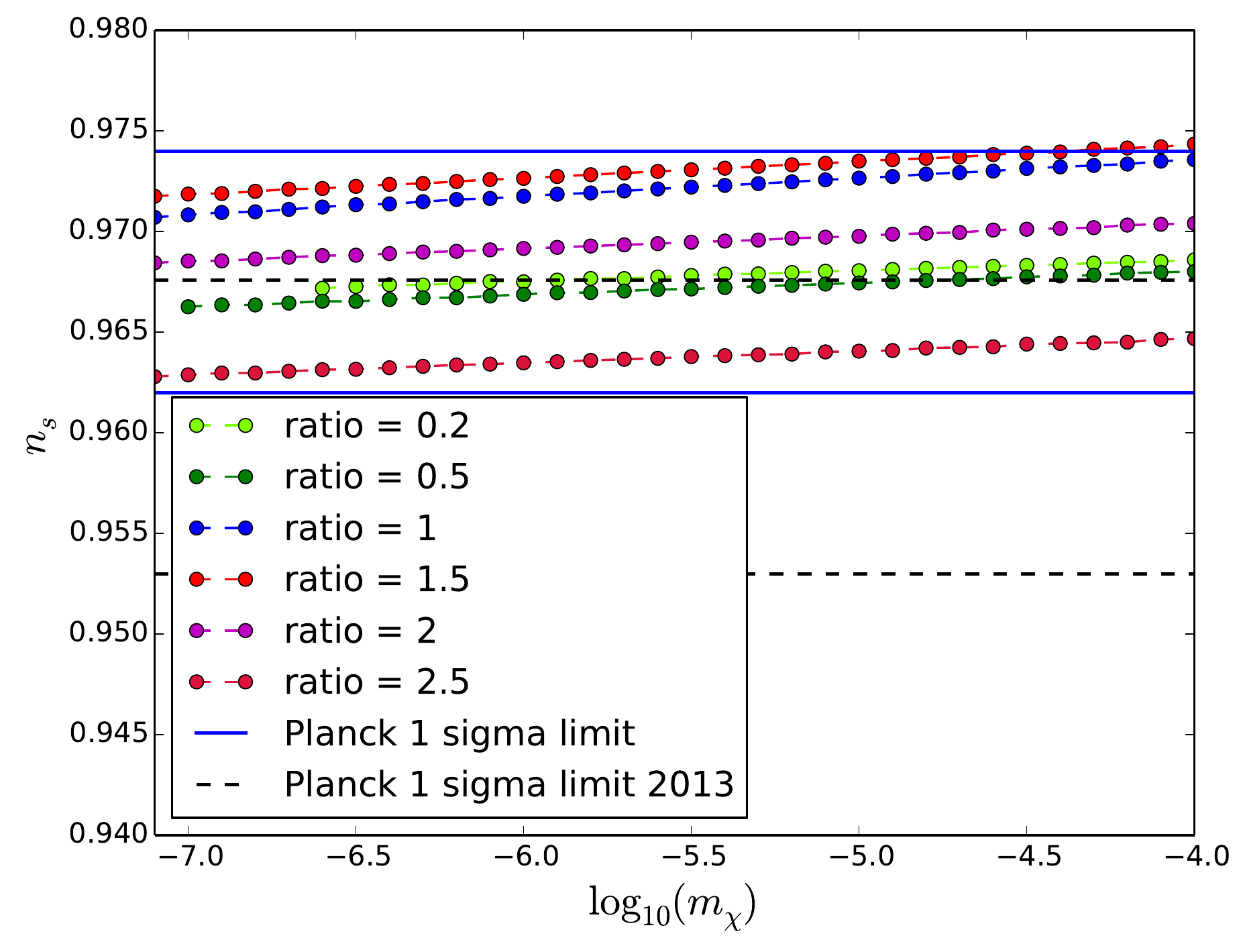}
\includegraphics[width=0.5\textwidth]{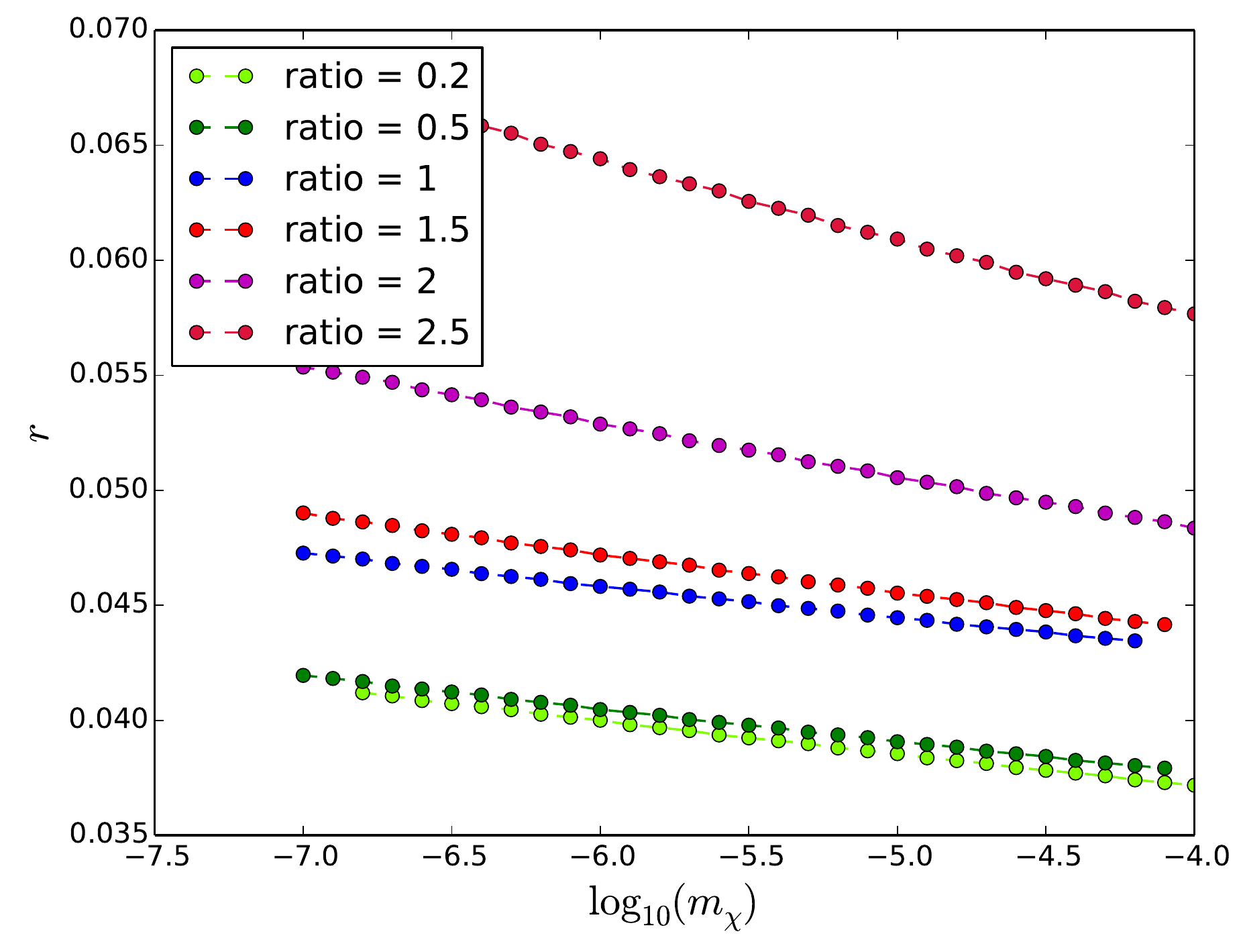}
\end{center}
\caption{Amplitude, spectral index and tensor-to-scalar ratio $r$ of the curvature perturbation power spectra at the pivot scale as a function of $m_\chi$ (in Planck units).}
\label{fig:mass-ratios-fields}
\end{figure}

The strongest constraint on the parameters  $m_\psi$ and $m_\chi$ comes from the amplitude $A_S$. In fig. \ref{fig:mass-ratios-fields} we show the amplitude $A_S$ at the pivot scale for different mass ratios $R_{m}= m_{\psi}/m_{\chi}$. As one can see, the observed amplitude of the primordial perturbations puts strong constraint on the masses for a given $R_m$, with narrower ranges for higher mass ratios. 

We also illustrate the results for the spectral index and the tensor--to--scalar ratio in fig. \ref{fig:mass-ratios-fields}. For comparison we plot the $1\sigma$ limit on $n_s$ coming from the Planck 2013 \cite{Ade:2013zuv} and Planck 2015 publications \cite{Planck:2015xua}. Interestingly, the predictions of the model are fully consistent with the 2015 $1\sigma$ results, but not with the 2013 $1\sigma$ results. When considering the $2\sigma$ limits, the discrepancy disappears. The constraints on  $n_s$ rule out mass ratios larger than 2.5 and  considering ratios below 0.2 restrains us to the regime of single field inflation driven by $\psi$ (see fig. \ref{fig:background-examples}). 
Under variations of the mass ratio $R_{m} \in [0.2, 2.5]$ the values predicted by this extension of the Starobinski inflation are in great agreement with current observational data.

The tensor--to--scalar ratio appears to increase with the mass ratio. Given the $1\sigma$ constraints on $n_s$, the tensor--to--scalar ratio is limited to be $r<0.07$, but higher values of $r$ are possible if the limits on $n_s$ are relaxed. 

\newpage
\clearpage
\section{Conclusions}
We considered the simplest extension to the Starobinsky model, where the quantum corrections to gravity and a scalar field drive inflation. Our analysis was done in the Jordan frame, where the theory looks like two scalar fields, one with canonical and the other with non-canonical kinetic terms, to second order in slow-roll parameters. The great agreement between the first and second order analysis confirms that a full integration of the equations is not necessary for our analysis. 

We show that there are regions of parameter space which yield realistic inflation without the need for fine-tunning (in the sense that for a given $m_\chi$ one can find an $m_\psi$ (or $\mu$) in the range considered here). We showed that the model preserves its appealing features; the spectral index is within the limits of the latest observational results, the amplitude of the perturbations at the chosen pivot scale is sufficiently small and the tensor--to--scalar ratio is $<0.1$.  

We embed our model in the framework proposed in \cite{Ellis:2014gxa}, but we later show that by allowing the fields to have different masses, you do not necessarily violate the latest constraints coming from the Planck satellite. If this model is to be embedded in Supergravity, the two fields don't necessarily need to belong to the same multiplet.

In future work we will be considering particle production at the end of inflation. We will analyse the model in the context of pre- and reheating to verify the feasibility of the simplest extension to Starobinsky inflation.

\vspace{0.5cm}

\noindent {\bf Acknowledgements:} CvdB is supported by the Lancaster-Manchester-Sheffield Consortium for Fundamental Physics under STFC grant ST/L000520/1.

\end{document}